# The impact of Egypt's accession to the BRICS group on the foreign exchange crisis in Egypt.

Yasmeen Fekery Yaseen El Khodary[1], Mousa  Gowfal Selmey Gowfal Selmey[2],  Elsayed farrag  Elsaid Mohamed Elsayed[1,3]

[1]Department of Economics, Faculty of Commerce, Damietta University, Egypt, [2]Department of Economics, Faculty of Commerce, Mansoura University, Egypt,[3]Faculty of law and Economics, Islamic University of Madinah, Saudi Arabia.

 *Email: yelkhodary@du.edu.eg

**Abstract**

   This paper investigated the impact of Egypt's accession to the BRICS bloc by studying the participation rate of BRICS countries in Egypt, as well as the exposure rate and the ratio of foreign investment of BRICS countries in Egypt to the total foreign investment in Egypt, as well as Egypt's export opportunities in the BRICS markets and the impact of these variables on the ratio of foreign assets in Egypt during the period from 2000 to 2024, In order to predict the impact of Egypt's accession to the BRICS bloc in the coming years as well. The study used the ARDL model to determine the impact of the four independent variables on the dependent variable (foreign assets), which is the ratio of foreign assets. It also used automatic ARIMA model to predict the values of the study variables during the period from 2024 to 2030.

   The study concluded that there is a negative relationship between net foreign assets and each of Egypt's export opportunities in the BRICS countries, the economic participation rate, and the ratio of foreign investments in the BRICS countries to the total foreign investments in Egypt. While there is a strong positive relationship between net foreign assets and the economic exposure rate. the independent variables are forecasted for the period from2024 to 2029 using automatic ARIMA model. Both the economic exploration rate, the economic participation rate, and Egypt's export opportunities in the BRICS countries decrease from (44.08-199.99-0.21) in 2025 to (37.30 -145.41 -0.15) in 2029. While the value of BRICS net foreign investment in Egypt increased from 70.33 in 2025 to 81.70 in 2029.

- **Introduction**

Egypt has been exposed in recent years to several political and economic crises and fluctuations, which had a major impact on many macroeconomic variables. Egypt has been exposed to a foreign exchange crisis, which is not new, but rather a result of the accumulation of major events that affected indirect investments in Egypt, bonds and treasury bills, including the Corona pandemic, then the Russian-Ukrainian war and the subsequent impact on supply chains. This contributed to a significant increase in inflation rates, and the international community tried to control it by raising interest rates. This had negative effects on emerging markets, including Egypt. (Mpofu, T.R. 2016).

Among the factors that led to the exacerbation of the foreign exchange crisis in Egypt are the spread of the black market and speculation on the dollar price and the occurrence of an unprecedented price fluctuation because of the importer's reliance on the dollar to meet the needs of the market. This led to an increase in the prices of production requirements and thus an increase in the dollar price in the black market to unreal rates due to speculation, exploitation and lack of control. (Ahmed, Y. et al,2020)

The large shortage of foreign exchange is due to the decline in Egypt's revenues from the dollar, the most prominent of which is the transfers of Egyptians abroad and the returns from tourism, as well as the recent decline in the revenues of the Suez Canal. The transfers of Egyptians abroad decreased during the past year by about 10 billion dollars, as it recorded 31.9 billion dollars in 2021/2022 compared to 22.1 billion in 2022/2023. The Suez Canal revenues also declined due to the Houthi attacks on some commercial ships in the Red Sea.

Egypt suffers from a structural problem related to net inflows of foreign currency entering and leaving the Egyptian economy. Over the past 50 years, Egypt has suffered from a doubling of the volume of imports compared to exports, and there is a severe shortage of foreign currency, which is compensated for by tourism, foreign direct investment, the Suez Canal, and other sources. The dollar in Egypt, as well as grants and loans, which achieves a balance and at times a surplus of currency in Egypt.( Elshahawany. D 2022)

Joining international blocs and organizations has become a feature of the era and a factor of strength that countries seek to achieve politically and economically. Despite Egypt's joining many international blocs, the size of the benefits granted to Egypt from this exchange is not commensurate with its economic status, in addition to the non-trade restrictions imposed on Egypt to dominate political decision-making. Hence, the importance of joining the BRICS bloc came about because it is an emerging bloc and includes many opportunities for emerging markets.

The main question of this paper is what the impact of Egypt's is joining the BRICS bloc on the foreign exchange crisis in Egypt. the paper also attempts to predict the impact of Egypt's joining the BRICS bloc on some economic variables such as the rate of economic participation, the rate of economic exposure, and the export opportunities available to Egypt in the BRICS countries over the next five years.

The study aims to analyze the impact of Egypt's partnership or accession to the BRICS group on the volume of foreign direct investment in Egypt, the volume of Egyptian exports, and the opportunities available to Egypt, as well as its impact on the exchange rate and the foreign exchange crisis, and study the volume of foreign investments of the BRICS countries in Egypt and their percentage of the total foreign investments in Egypt, through the following sub-objectives:

- Analyzing the volume of exports and imports of the BRICS countries and studying Egypt's export opportunities in the BRICS countries with the aim of working to activate and enhance means of cooperation in a way that achieves the consolidation of economic relations and expands the scope of the volume of trade exchange between the Egyptian market and the markets of the BRICS countries. (Ashraf, K. 2024)
- The study also aims to predict the impact of Egypt's accession to the BRICS bloc on the volume of foreign investments of BRICS countries in Egypt, the rate of economic participation, and the rate of economic exposure of BRICS countries during the five years following Egypt's accession to the BRICS bloc.

The study will be divided into the following parts:

1. **Introduction**

2. **literature review:**
3. **Analysis of the main causes of the foreign exchange crisis**
4. **The BRICS block :general objectives and its development.**
5. **Evaluation of the impact of Egypt's accession to BRICS on the foreign exchange market.**
6. **Appendix**

**2. literature review:**

By reviewing previous literature and studies about groupings and blocs Economic, especially the BRICS group, the study found some studies related to the research, including the research by **Attia Masoud, (2018)** This study aimed to identify the development of the BRICS bloc, and to demonstrate the economic position of the BRICS bloc. Among the countries of the world. The future of the BRICS bloc, the role it is expected to play in the global system, and the challenges facing the bloc. The study concluded that the BRICS bloc is developing significantly from being a political bloc to a political-economic bloc that seeks to deepen economic ties between member states and play a major role in the global financial and economic system.

In addition to the bloc's contribution to global output as well as to global trade in goods and services, and the bloc has now occupied a position Major in the world in terms of direct foreign investment flows and providing development aid to developing countries according to a model different from the model followed by traditional donor countries, the BRICS bloc is distinguished by the fact that it is based on mutual benefit and not conditionality. The study recommended the necessity of expanding the bloc's membership by including new countries, as well as expanding the conclusion of economic agreements between the bloc and other active blocs on the global scene.

Another Study by **Bas Hooijmaaijers, (2021)** examined how and why China and the BRICS are reshaping global economic governance, and to what degree the BRICS and BRICS institutions represent anything new. More importantly, it analyzed China's use of the BRICS as a tool for its objectives in reshaping global economic governance and the potential for its independent initiatives to undermine BRICS' impact on global economic governance. Because of its politico economic weight, China exerts great weight on the BRICS exceeding that of its partners.

This study found that China is not only reshaping the existing global economic order by becoming another voice and actor in its operations but is also launching new economic governance institutions both by itself as well as in the BRICS format. The bottom line is that the BRICS are a multilateral group with no grand vision that has identified several fields of cooperation. However, what is critical is the domestic political economy and interests of China, India and the other BRICS countries that for instance driven by their interests all hold diverging positions and preferences in the international system.

The Study by (Kareem Ashraf 2024) presents an ex-ante impact assessment of a hypothetical FTA between Egypt and the BRICS from an Egyptian economy-wide and sectorial perspectives, with a granular look into manufacturing. the chosen methodology is a static SAM-based Computable General Equilibrium model calibrated to Egypt's 2018-2019 Social Accounting Matrix (SAM). With respect to existing literature, the paper uniquely stands in considering an Egypt-BRICS FTA with a granular assessment of manufacturing subsectors and in running a simulation of Egypt's trade liberalization with the wider BRICS alliance, including the accession of Saudi Arabia, UAE, Ethiopia, and Iran, which joined the bloc along with Egypt in January 2024.

Besides the wider BRICS simulation, the model is used to run a simulation with core BRICS members. Magnified upon considering the wider block, results predict an increase in real consumption across all household income quantiles with the poor generally reaping more of the welfare gains, defined as the increase in household real consumption. Real GDP expands while inflation is imported on the back of local currency depreciation, implying a positive exchange rate pass-through.

The **study by Robert Greene (2023)** presented a set of important points, including the role of the dollar in emerging markets and the most important concerns facing BRICS members regarding the macroeconomy, and how geopolitical concerns drive it, and touched on the idea of the rise of the renminbi as a common currency for the BRICS group and the use of other BRICS currencies in cross-border trade, as well as India's emerging efforts to develop the use of Rupee.

The study found there Another area of uncertainty that will influence the growth in global use of BRICS currencies—particularly the renminbi—is how Washington responds to the range of policy initiatives aimed at expanding the use of non-dollar financial channels. Some of these channels are increasingly being used to facilitate transactions with regimes and entities heavily sanctioned by the U.S. government, as evidenced by a 2023 remark by the head of Iran's central bank that "the financial channel between Iran and the world is being restored." Complicating any U.S. policy responses is the reality that strategic partners like India are involved in efforts to build out and use such channels. Also, it seems unlikely that U.S. President Joe Biden's administration will take actions that meaningfully target Chinese institutions' ability to facilitate renminbi trade payments to and from Russia; without the ability to pay Russian firms in dollars, many U.S. strategic partners appear to be using these channels.

Additionally, the Study by **H. Zhao, D. Lesage (2020)** examines and explains the outreach activity of the BRICS group of Brazil, Russia, India, China and South Africa. We focus on two research puzzles: a) the motivations and b) the form and degree of the institutionalization of BRICS outreach. We define outreach as collaborative interaction among BRICS and other actors within and outside the BRICS area and focus on outreach to governments of non-BRICS countries and national top officials representing regional organizations. First, we offer a theoretical framework based on the experiences of the Group of 7/8 (G7/8) and the Group of 20 (G20), considering both commonalities and differences with BRICS. Second, we provide a detailed empirical analysis of BRICS outreach over time. Third, we explain BRICS outreach considering the theoretical framework and enrich it based on our findings. Methodologically, we draw empirical information from official documents, news media and academic literature.

We argue that the outreach activity of a major power grouping is reflective of its internal cohesion, as well as of how it defines its own position in the world and how it is perceived by others. This research offers a timely contribution to the ongoing debate on BRICS and the under-researched BRICS outreach process as a part of the overall institutionalization of BRICS.

Another study by **Elshahawany. D, (2022)** aimed to addressee the impact of exchange rate volatility on economic growth in Egypt. The empirical analysis is based on data spanning the period 1992–2018. Exchange rate volatility calculation is based on the Generalized Autoregressive Conditional Heteroskedasticity (GARCH). Using the corresponding Mixed Data Sampling (MIDAS) regression models, we found that exchange rate volatility has a significant negative impact on economic growth in Egypt.

Additionally, the Study by **Jawadi F. Sushanta K. Mallick, Ricardo M. Sousa (2016)** Using a Panel Vector Auto-Regressive (PVAR) model, we assess the macroeconomic impact of fiscal policy and monetary policy shocks for five key emerging market economies—Brazil, Russia, India, China and South Africa (BRICS).

We show that monetary contractions lead to a fall in real economic activity and tighten liquidity market conditions, while government spending shocks have strong Keynesian effects. We also find evidence supporting the existence of accommodation between fiscal policy and monetary policy, which is crucial for economic and political decision-making. Our results are robust even after controlling for periods of extreme instability, such as economic and financial crises.

Another study by **Alabdulwahab S. Abou-Zaid A. (2023)** aims to empirically investigate the sources of real exchange rate fluctuations in Egypt using structural vector autoregression (SVAR). The data covers the period between 1980 and 2016, where the exchange regime has been changed more than once.

This paper investigates the source of real exchange rate fluctuations for the period between 1980 and 2016 using the SVAR method. The SVAR method will incorporate real gross domestic product (GDP), real effective exchange rate (REER) and price level in a multidimensional equations system. However, impulse response function (IRF) and error variance decompositions (EVDC) will be generated by the system to have a behavioral insight of real exchange rate in response to economic shocks.
The IRF and EVDC results indicate a significant impact of demand shocks over the real exchange rate relative to supply shocks and monetary shocks in the period between 1980 and 2016. On the other hand, monetary shocks will have a negligible effect on the real exchange rate in the short run and converging to its previous level in the covering period of the study.

## 3. Analysis of the main reasons for the foreign exchange crisis

Egypt experienced problems with its official currency regime in late 2012. Since 2003/2004, the stability of the exchange rate regime had depended on the ability of the Central Bank to accumulate large foreign reserves that would cover 9-10 months of annual imports. The Central Bank could use these reserves to meet the demand for dollars needed to pay for imported goods, and thus effectively defend the value of the pound. Egypt is a major importer of food and fuel, and therefore managing the exchange rate is important to control inflation, especially the price of primary products.( International Monetary Fund, 2011).

During 2011 and the political turmoil that followed, growth rates and foreign exchange reserves declined significantly. This was due to Egypt suffering from capital shortages and a sharp decline in foreign direct investment and tourism. Between 2011 and 2012, the Central Bank used its reserves to defend the value of the pound by financing the import of basic food products and fuel by state agencies.

By December 2012, foreign reserves had shrunk by $35 billion compared to January 2011 to $15 billion, reducing the central bank's ability to meet demand for dollars. (Brahim, M., Nefzi, N. and Sambo, H. 2018)

However, the largely politically motivated expansion of foreign loans provided aid in the form of long-term deposits in the central bank and artificially kept the regime alive. In July 2013, Egypt received about $8 billion, mostly from Qatar and, to a lesser extent, Turkey. After 2013, these special relationships abruptly collapsed (many Qatari deposits and loans were returned), to be replaced by greater support from Saudi Arabia, the United Arab Emirates and, to a lesser extent, Kuwait (Bushkov, T. 2019).

During the period 2011–2016, Egypt received a total of $29 billion in aid, cheap loans, long-term deposits in the central bank and free shipments of oil and gas. However, the administration's lack of economic vision and unwillingness to cut costs meant that the missed opportunity of these unprecedented inflows was lost. Until the support of the Gulf Cooperation Council countries came, which could help mitigate the deflationary effects of saving and facilitate the implementation of reforms to address long-standing structural weaknesses in Egypt's finances. (Boshkov ,T., op. cit.).

Recognizing the volatility that could come with rising prices and the dismissal of government employees, the new leadership has chosen to use these funds to implement reforms. Regarding the foreign exchange market, the weighted average exchange rate of the dollar in the interbank market reached EGP 15.6818 by the end of June 2021, compared to EGP 16.1384 by the end of June 2020, with the pound's value increasing by 2.9% during the fiscal year.

The weighted average exchange rate of the dollar reached EGP 16.7057 in the interbank market by the end of June 2019, with the pound's value increasing by 3.5% during the fiscal year 2019/20. At the end of December 2020, the weighted average exchange rate of the dollar reached EGP 15.7321 at the end of December 2020, with the pound's value increasing by 2.6% during the July/December period of the fiscal year 2020/2021. As for the net international reserves at the Central Bank, they increased by about $2.4 billion, at a rate of 6.3% during the fiscal year 2020/2021, to reach about $40.6 billion by the end of June 2021, covering about 6.9 months of commodity imports. (Alabdulwahab, S.& Abou-Zaid, A.2023)

As for the net international reserves at the Central Bank, they decreased by about $6.3 billion, at a rate of 14.2% during the fiscal year 2019/2020, to reach about $38.2 billion by the end of June 2020, covering about 7.3 months of commodity imports.

When a devaluation is expected on the market, downward pressure on the currency can only be settled (neutralized) by increasing the interest rate. To increase the rate, the Central Bank must reduce the money supply, which in turn will cause an increase in demand for the currency. The bank can do this by selling foreign reserves to create an outflow of capital. When a bank sells part of its foreign exchange reserves, it receives payment in the form of a domestic currency, which keeps it outside the turnover as a means. (Elshahawany. D, op,cit)

Also, the reliance on the exchange rate cannot last forever, both in terms of the reduction of foreign reserves, as well as economic and political factors, such as the increase in unemployment. Devaluation of the currency through a fixed exchange rate makes domestic goods less expensive than foreign goods, which increases demand for workers and output, and gives a competitive advantage to domestic exports in the case of producing countries. (Keefe, H.G. 2014)

In the short run, devaluation also raises interest rates, which must be settled by the Central Bank by increasing the money supply and increasing foreign reserves. At the end of 2016, Egypt has the highest amount of foreign exchange reserves of $ 24.26 billion in the last 5 years. The lowest amount of foreign exchange reserves was in June 2013 with the amount of 14.94 billion. ( Bassiouny, A., & Tooma, E. 2018).

The exchange rate of the US dollar has developed significantly in recent years in Egypt. The table (1) shows the development of the Egyptian pound exchange rate over the past ten years, as it took an upward trend, and the exchange rate increased from 7.14 in 2014 to 48.06 in 2024.

This shows the decline in the value of the Egyptian pound and the deterioration of its value, which led to higher inflation rates, a decrease in the purchasing power of consumers, and an increase in the cost of imports, which leads to an increase in demand for the US dollar, which increases the balance of payments deficit, reduces consumer spending by individuals, and reduces the rate of economic growth.

## 4. The BRICS bloc: general objectives and its development.

The concept of economic blocs has emerged and developed in industrialized countries and has become important and necessary in the international community. Economic blocs are the gathering of a few countries that express a degree of economic integration, in the form of a customs union or a free trade zone and other interests that bring those countries together. Joint efforts to maximize mutual benefits, achieve the greatest returns, and achieve high levels of economic well-being for people. (Mlachila, M. and M. Takebe, 2011).

BRICS was established by the union of the following four countries, namely (Brazil, Russia, India, and China), which are countries with emerging economies that witness high growth rates and a large population, despite the political differences, as the political systems differ, as well as the difference in history, culture, and geography. (O'Neill. J., 2007).

In 2006, the foreign ministers of the four countries met on the sidelines of the United Nations General Assembly in New York, and in 2009 the inaugural summit of the BRICS group was held in Yekaterinburg, Russia. Since that time, summits have been held annually, and the summit adopted the group's joint statement and a joint statement on Global food security. (Mwase, N., 2011)

The second summit took place in Brasilia, Brazil in 2010. A joint statement was issued after the summit, and a memorandum of cooperation was signed. In 2011, the third summit was held in Sanya, China, and the summit's theme was "Broad Vision and Shared Prosperity." In 2012, the fourth summit was held in New Delhi, India, under the title "BRICS Partnership for Global Stability and Security." At the conclusion of the summit, a declaration was issued defining the common positions of the BRICS countries. On global issues and giving a road map to enhance cooperation among the BRICS countries. (Branco R. May 2015)

The fifth summit, which was the last in the first session of the summits, was held in South Africa under the slogan "BRICS and Africa: Partnership for Development, Integration and Industrialization."

The BRICS group represents about $56.65 trillion of global GDP, or 26% of the global economy. Member states have a total cash reserve of about $4 trillion and contribute 16% of global exports and 15% of global imports of goods and services.

The member states seek to use a unified currency for their trade exchanges, and the BRICS countries also encourage the use of local currencies among them and have already begun to do so. (Mwase, N & Yang, Y. March 2012)

Its membership includes five countries with a population of about 2.88 billion people, equivalent to 41% of the world's population, and the area of the member states covers about 40 million square kilometers, equivalent to approximately 26% of the total area of the world. Member states established the New Development Bank (NDB) in 2014, and some see it as an alternative to the World Bank and the International Monetary Fund.

in September 2022, the Russian Foreign Ministry announced that there were 15 countries interested in joining BRICS. In October 2022, China announced its support for expanding the alliance's membership. During talks in South Africa on June 1 and 2, 2023, the foreign ministers of the BRICS countries announced that the group is open to the joining of new members, as part of the group's demand to restore balance in the global system. (Greene, R. December2023)

In 2023, the 15th summit was held in Johannesburg, South Africa, from August 22 to 24. Saudi Arabia, the UAE, the Comoros, Gabon, Iran, Cuba, the Democratic Republic of Congo, and Kazakhstan have all sent representatives to Cape Town for talks about possible membership, and Egypt, Argentina, Bangladesh, Guinea Bissau, and Indonesia have already participated.

Many developing countries and emerging markets have expressed their interest in joining the group, and recent reports indicate that the countries that are ready to join BRICS are (Thailand, Venezuela, Indonesia, Bangladesh, Pakistan, Kazakhstan, Nigeria, Zimbabwe, Turkey, the Kingdom of Saudi Arabia). Saudi Arabia, Egypt, Bahrain, Algeria, Sudan, Syria, UAE, Tunisia, Senegal, Argentina, Belarus, Mexico, Uruguay).

On August 24, 2023, the BRICS leaders announced the accession of 6 new countries, namely (Egypt, Saudi Arabia, the Emirates, Argentina, Ethiopia, and Iran), as of January 1, 2024, and will be called "BRICS Plus" or "BRICS+." Calls to expand the group dominated the agenda of its three-day summit in Johannesburg, South Africa, and also revealed divisions between countries over criteria for accepting members. South African President Ramafo said the group had agreed on guidelines for the BRICS expansion process.

Objectives of the BRICS grouping: The BRICS grouping of countries aims to adopt an alternative vision for the international economic system, in order to serve the interests of developing countries in a way that brings them benefits and avoids the deterioration of their economic and political conditions, especially in the event of various international crises, by striving to achieve the following goals: (Mwase, N. and Y. Yang, 2011)

The BRICS grouping aims to restore international economic balance and end the unipolar policy and the United States' hegemony over global financial policies, by changing the global economic and financial system represented by the institutions of the Bretton Woods system such as the Monetary Fund, the World Bank, and the World Trade Organization, and introducing quota reforms. (Aydın, L, 2016).

Voting rights for those institutions and finding an effective and real alternative to them, and improving the representation of the membership of the member states in the countries of the group in the United Nations institutions, the G20 and others, in addition to searching for ways to strengthen their negotiating positions in the process of forming a new world order, in addition to achieving economic and political integration between them Community countries. (Castel, V., Mejia. P, Kolster. J,2011)

- Developing the infrastructure in the countries of the group and not resorting to Western institutions in times of economic crises by finding an effective way to grant and exchange loans between the countries of the group, and strengthening the global economic safety network for those countries and sparing them the pressures of loans and their interest from Western institutions. (Fredj J, et al, 2016).
- The BRICS grouping aims to adopt a unified position to fight poverty and move towards development. Sustainable, and rejecting Western hegemony over international institutions, whether financial, economic, political, or social, and over international decisions. (Larionova, M.2016)
- The BRICS grouping aims to reject the protectionist trade policies represented by America imposing taxes and fees on imports of goods and services from developing countries, including the group's countries, which the United States has begun taking steps to implement. (Samake, I. and Y. Yang, 2011)

- The BRICS group seeks to raise the level of trade exchange between members, as well as raise the level of trade exchange with developing and poor countries such as the African continent. (Fred. J et al, 2014)

The BRICS grouping consists of five countries from four different continents. They are considered different Significantly different from other forms of groupings, alliances and organizations that the international arena has witnessed before, there is no specific common link between the five countries, whether political, economic, cultural or otherwise, nor are they linked by a geographical or regional scope, and there is a clear difference in their degrees of economic development and levels. Productivity, and even the political positions among them, are relatively different, because they include countries that differ greatly in political orientations and economic systems and represent different global trends. (Hooijmaaijers ,B. 2021).

Despite all these differences, these five countries share an important link, which is that they do not belong to the circle of Western civilization, as they constitute several different civilizations. The most important link that brought these five countries together is the political link, on the basis of which this group arose, which is represented by the rejection of Western hegemony over the economy and global politics, which has caused the global financial crises that most countries of the world are suffering from, and these countries have in common that they are characterized by with rapid economic growth and its quest to reach a distinguished economic position in the world.( Zhao H., Lesage D.2020)

BRICS countries are keen to develop greater independence from the Western-led international monetary system. Approximately 90% of global foreign exchange transactions are conducted in dollars and flow primarily through US and European banks. Western financial sanctions on Russia underscored the powerful systemic influence the US still holds through Bretton Woods institutions and its central role in the global financial system. Because BRICS includes leading commodity exporters and importers, however, the group can become a conduit for foreign exchange transactions in currencies other than US dollars. (Shelepov., A, 2015).

The NDB, for instance, has issued about one-fifth of its loans in Chinese yuan. Russia, China, and other BRICS members also aim to promote digital currencies.

The group has launched the beta version of a payment app—BRICS pay—that enables transactions in several non-dollar currencies. That could help nations ease their reliance on the US and make them less vulnerable to sanctions and foreign exchange volatility during financial crises. From a governance perspective, BRICS has established the Payment Task Force, the Think Tank Network for Finance, and the Contingent Reserve Arrangement, establishing a pool of reserves that can be used in place of IMF funds to help nations address financial crises.

> BRICS could make a significant global impact in the following five areas:

**Energy**: BRICS brings together both some of the world's biggest energy producers and buyers. With the addition of Iran, Saudi Arabia, and the UAE, BRICS member states account for around 32% of world output of natural gas and 43% of crude oil. If Kazakhstan, Kuwait, and Bahrain are admitted, those shares will rise further. BRICS nations also account for 38% of global petroleum imports, led by China and India. If all new applicants are admitted, that would rise to 55%.

During times of volatility in energy markets, having many of the biggest energy buyers and sellers within the same group could give rise to a parallel energy trading system. That would allow for transactions among BRICS and economies outside the Western-led financial system and potential future sanction programs, and it would perhaps give them the ability to influence oil prices. (Larionova. M., Shelepov., A, 2016)

**Trade networks**: Trade has been a major driver of the economic development of BRICS. The share of global trade in goods transacted among the group's current members more than doubled, to 40%, from 2002 through 2022. This trend becomes clearer when looking at the increasing dependence of specific BRICS economies on trade with fellow BRICS members. China's growing role as a supplier of industrial and consumer goods, as well as an importer of commodities, has been a key force for integration. China has become a major market for Brazilian soybeans and iron ore, for example, and a major exporter of advanced goods such as electric vehicles, solar panels, and heavy machinery. Western sanctions relating to the war in Ukraine, moreover, have led to the diversion of Russian exports to BRICS markets, notably China and India.

Although a handful of BRICS members have free trade agreements (FTAs) with each other through blocs such as the Gulf Cooperation Council and Pan-Arab Free Trade Area, there is currently no FTA covering the entire ten-nation group. India withdrew mid-negotiation from Asia's Regional Comprehensive Economic Partnership, which includes China. BRICS could, however, serve as a forum for widening intra-BRICS market access in various ways. It has already convened a Digital Economy Working Group, for example, and has established a framework for promoting cooperation in professional and business services trade. (Correa, Luiz et al, 2014).

**Infrastructure and development financing**: The greatest progress so far in BRICS institution building has been in project and development finance. The New Development Bank (NDB), capitalized at $100 billion, largely complements China's Belt & Road initiative. Egypt, India, Russia, Saudi Arabia, and UAE are also shareholders in the China-led Asian Infrastructure & Investment Bank (AIIB) and have received loans from it. By 2023, the NBD and AIIB combined had committed more than $71 billion in credit across a range of sectors, including infrastructure, public health, and clean energy. Such projects generate significant revenue for BRICS companies. The addition of Saudi Arabia and other cash-rich economies, moreover, could expand and diversify the financial resources of BRICS. ( Azevedo. D, et al,2024).

**Monetary policy**: BRICS countries are keen to develop greater independence from the Western-led international monetary system. Approximately 90% of global foreign exchange transactions are conducted in dollars and flow primarily through US and European banks. Western financial sanctions on Russia underscored the powerful systemic influence the US still holds through Bretton Woods institutions and its central role in the global financial system. Because BRICS includes leading commodity exporters and importers. (Ozcelebi, O. 2019).

However, the group can become a conduit for foreign exchange transactions in currencies other than US dollars. The NDB, for instance, has issued about one-fifth of its loans in Chinese yuan. Russia, China, and other BRICS members also aim to promote digital currencies. ( Azevedo. D, op. cit)

The group has launched the beta version of a payment app—BRICS pay—that enables transactions in several non-dollar currencies. That could help nations ease reliance on the US and make them less vulnerable to sanctions and foreign exchange volatility during financial crises. From a governance perspective, BRICS has established the Payment Task Force, the Think Tank Network for Finance, and the Contingent Reserve Arrangement, establishing a pool of reserves that can be used in place of IMF funds to help nations address financial crises. (Hongmei Li, Ruizhe Xu.,2023).

Technological cooperation. Space is an overlooked dimension of BRICS collaboration. There is a BRICS Space Cooperation Joint Committee, supported by longstanding partnerships between Russia and China and China and Brazil. BRICS has also established a Partnership on New Industrial Revolution and a Center for Industrial Competencies. These initiatives aim to spur cooperation and innovation in leading-edge technologies in areas such as intelligent manufacturing, artificial intelligence, digitization, and clean energy. The efforts could help more emerging markets get in on the ground of new technologies, improve their capacity to create intellectual property, and adopt alternative technical standards. (Marc J, et al,2024).

## 5. Evaluation of the impact of Egypt's accession to BRICS on the foreign exchange market

- **Methodology**

The main aim is to investigate the long run and short run relationships between net foreign assets and other determinants. This can be done using cointegration analysis and error correction model. But Cointegration test and error correction model are used within the an ARDL framework because Johansen cointegration test cannot be applied directly if the variables of interest is not all I(1). That is, it is not applicable for mixed order of integration for the variables of interest or all of them are not non-stationary. So, an alternative method is needed if the variables are of mixed orders, or some of them are non-stationary, this method is ARDL model.

An autoregressive distributed lag (ARDL) model is an ordinary least square (OLS) based model which can be used if the models are of mixed orders or some of them are non-stationary. This model takes sufficient numbers of lags to capture the data generating process in a general-to-specific modeling framework.

Using a simple linear transformation, a dynamic error correction model (ECM) can be derived from ARDL. Also, the ECM integrates the short-run dynamics with the long-run equilibrium without losing long-run information and avoids problems such as spurious relationship resulting from non-stationary time series data. (Shrestha, M.B. and Bhatta, G.R., 2018)

To illustrate the ARDL modeling approach, the following simple model can be considered:

$$y_t = \alpha + \beta x_t + \delta z_t + e_t$$

The error correction version of the ARDL model is given by

$$\Delta y_t = \alpha_0 + \sum_{i=1}^{p} \beta_i \Delta y_{t-i} + \sum_{i=1}^{p} \delta_i \Delta x_{t-i} + \sum_{i=1}^{p} \gamma_i \Delta z_{t-i} + \lambda_1 y_{t-1} + \lambda_2 x_{t-1} + \lambda_3 z_{t-1} + u_t$$

The first part of the equation with $\beta$, $\delta$ and $\delta$ represents short run dynamics of the model. The second part with $\lambda_s$ represents long run relationships. The null hypothesis in the equation is $\lambda_1 + \lambda_2 + \lambda_3 = 0$, which means non-existence of long run relationship. In our study we have 3 dependents which are closed price of FTSE-100, USD/GDP, EUR/GDP, and only 1 independent variable which is sentiment score that calculated from tweets.

- **Study variables**
    - NFA: Net foreign assets
    - EPR: Economic participation ratio
    - EER: Economic exposure ratio
    - EoF: Export opportunities for Egypt in the BRICS countries
    - PFI: Percentage of foreign investments from BRICS in Egypt/total foreign investments in Egypt (Sebastian, R., 2008).

**Cointegration Results**

Results of the bounds test procedure for co-integration analysis between net foreign assets and their determinants are presented in the table below.

From the table (4) it is clear that, the f-calculated is lower than the lower bound for each significance levels, which mean that at 95% confident level the null hypothesis "no long run relationship exist" is accepted, this means that there is no cointegration relationship (i.e., long run relation) exists between net foreign assets and its determinants.

- **Results of the ARDL Model**

The ARDL model was estimated to be based on the Akaike Information Criterion (AIC) from the table (5), we can conclude that

- EOF has a negative significant coefficient at 95% confidence level, this means that every increase in EOF by 1 unit the net foreign assets will decrease by 14.97 units this with fixing all other factors.

- EPR has a negative significant coefficient at 95% confidence level, this means that every increase in EPR by 1 unit the net foreign assets will decrease by 3.12 units this with fixing all other factors.

- EER has a positive significant coefficient at 95% confidence level, this means that every increase in EER by 1 unit the net foreign assets will increase by 11.39 units this with fixing all other factors.

- PFI has a negative significant coefficient at 95% confidence level, this means that every increase in PFI by 1 unit the net foreign assets will decrease by 5.119 units while fixing all other factors.

- **Comparing the effect of variables**:

First, we will compare the coefficients of the independent variables using linear combination hypothesis testing, the following table presents the labels of each coefficient

Coefficient Labels
Equation: EQ01
Method: ARDL

| Variable | Coefficient |
|---|---|
| EOF | C (2) |
| EPR | C (3) |
| EER | C (4) |
| PFI | C (5) |

we will test the following hypothesis

$$H_0: C2 = C3 = C4 = C5$$

The results are presented in the following table, and from it we can conclude that we will reject that all coefficients are equal which means that coefficients are statistically significant differently from each other.

| Test Statistic | Value | df | Probability |
|---|---|---|---|
| F-statistic | 10.893178 | (3, 16) | 0.000 |
| Chi-square | 15.679535 | 3 | 0.000 |

From the following table it is also clear that each pair of coefficients is statistically significant differently from each other,

| Normalized Restriction (= 0) | Value | P-value |
|---|---|---|
| C(2) - C(5) | -14.850 | 0.0000 |
| C(3) - C(5) | 1.993061 | 0.0487 |
| C(4) - C(5) | 16.51737 | 0.0000 |
| C(2) - C(3) | -14.843 | 0.0000 |
| C(2) - C(4) | -14.263 | 0.0000 |
| C(3) - C(4) | -14.524 | 0.0000 |

From the following table and the values of the standardized coefficients the variable with the highest effect on the net foreign assets is EER, while the variable with the lowest effect is EOF.

| Variables | Standardized Coefficient |
|---|---|
| EOF | -0.1776 |
| EPR | -0.17803 |
| EER | 0.659788 |
| PFI | -0.20056 |

It is clear from the previous table that the economic exposure ratio is the variable that has the greatest impact on net foreign assets (0.659788) and is the variable that has a direct relationship with net foreign assets, compared to other variables that have an inverse relationship with net foreign assets (0,55619). Therefore, the strongest impact in the end is the direct relationship between the economic exposure ratio and net foreign assets.

- **Goodness of Fit**
  From the following tables, there is no serial correlation as Durbin Watson value is near to 2, also from Q-statistics probabilities, there is no serial correlation as p-value is greater than 0.05, also supported by graph as the residuals are scattered randomly, in addition the fitted value is almost the same as the actual values. Also, from the value of R-square, the estimated model could explain around 74% of the variation in the net foreign assets.

From the graph (3) residuals of the model follow normal distribution

From the table (8), there is no heteroskedasticity problem with confident 95% as the P-Value of Breusch-Pagan-Godfrey is greater than 0.05.

- **Forecasted values of the net foreign assets**

The study attempts to predict the values of the basic variables of the study in order to predict the impact of Egypt's accession to the BRICS bloc on the values of some variables, namely the rate of economic participation, the rate of economic exposure, and Egypt's export opportunities in the BRICS countries, and the rate of foreign investments of the BRICS countries in Egypt relative to the total foreign investments in order to predict the impact of Egypt's accession on these variables and their impact on the rate of net foreign assets in Egypt.

The aim is to forecast the values of the net foreign assets, as a **first step** the independent variables are forecasted for the period from 2024 to 2029 using automatic ARIMA[1] in E-views. The table (9) shows the forecasted values for all variables. Both the economic exploration rate, the economic participation rate, and Egypt's export opportunities in the BRICS countries decrease from (44.08-199.99-0.21) in 2025 to (37.30 -145.41 -0.15) in 2029. While the value of BRICS net foreign investment in Egypt increased from 70.33 in 2025 to 81.70 in 2029.

**Results and discussion:**

- The previous analysis of the foreign exchange market position in Egypt shows that Egypt has been suffering from significant and recurring fluctuations in the Egyptian economy during the recent period, which is evident in the rise in the value of the dollar against the pound, as it has risen by a very large percentage during the last ten years.
- It is clear from the standard model that there is a negative relationship between net foreign assets and each of Egypt's export opportunities in the BRICS countries, the economic participation rate, and the ratio of foreign investments in the BRICS countries to total foreign investments in Egypt. While there is a strong positive relationship between net foreign assets and the economic exposure rate, meaning that every increase in the economic exposure rate by one unit leads to an increase in net foreign assets by 11.39 units.
- The study expects that there will be an increase in net foreign assets and an increase in the ratio of foreign investments of the BRICS countries in Egypt, which is consistent with Egypt's trend towards increasing integration with the BRICS countries and thus increasing mutual investments as well as opening markets in both directions. This will have a positive and noticeable impact on the exchange rate of the Egyptian pound due to the BRICS bloc's tendency to deal in a new currency other than the dollar.

---

[1] An autoregressive integrated moving average, or ARIMA, is a statistical analysis model that uses time-series data to better understand the data set or predict future trends. A statistical model is autoregressive if it predicts future values based on past values.

7. **Appendix**

Table (1): the exchange rate of the Egyptian pound against the US dollar in a few years

| Years | Average exchange rates |
|---|---|
| 2014 | 7.14 |
| 2015 | 7.53 |
| 2016 | 8.88 |
| 2017 | 18.139 |
| 2018 | 17.8878 |
| 2019 | 16.7057 |
| 2020 | 16.138 |
| 2021 | 15.6818 |
| 2022 | 19.0084 |
| 2023 | 30.958 |
| 2024 | 48.06 |

Annual report of the Central Bank of Egypt, different years

Table (2): Descriptive statistics of variables

|  | NFA | EPR | EOF | EER | PFI |
|---|---|---|---|---|---|
| Mean | 57.07708 | 84.71052 | 0.370172 | 35.70168 | 23.80254 |
| Median | 114.6150 | 95.73162 | 0.321544 | 34.59471 | 20.86921 |
| Maximum | 348.5000 | 222.3054 | 0.799356 | 52.37719 | 76.69240 |
| Minimum | -875.2100 | -102.6220 | 0.190525 | 16.73387 | 0.000000 |
| Std. Dev. | 268.1170 | 75.28272 | 0.147376 | 8.488383 | 21.57623 |
| Skewness | -2.158404 | -0.448572 | 1.070325 | 0.143043 | 0.997260 |
| Kurtosis | 7.773563 | 3.122256 | 3.928802 | 2.960897 | 3.370090 |
| Jarque-Bera | 41.42174 | 0.819814 | 5.445053 | 0.083374 | 4.115076 |
| Probability | 0.000000 | 0.663712 | 0.065709 | 0.959170 | 0.127768 |
| Observations | 24 | 24 | 24 | 24 | 24 |

- **Descriptive Statistics**

Table (2) presents the descriptive statistics for the research variables. The results show the descriptive of the variables like mean, median as measures of central location, also standard deviation, minimum and maximum are presented. From the Jarque-Bera test, all variables are normally distributed with a confident 95%, as the p-value of the test is larger than 5%, except for the net foreign assets it is not normally distributed.

Also, from the plot it is clear that

- **NFA** has been fluctuating ups and downs over the years.
- **EPR** has almost increased over years.
- **EER** has been fluctuating ups and downs over the years.
- **EOF** has been fluctuating ups and downs over the years.
- **PFI** has been fluctuating ups and downs over years.

- **Empirical results**

As an initial step of the time series analysis, it is to validate the stationarity assumption. Stationarity assumption is tested using the Augmented Dickey-Fuller (ADF) test. The ADF test is one of the cited unit root tests in literature and commonly used. The ADF test is applied to determine whether the data series is stationary (has no unit root) or not, by calculating the respective statistics and p-values in the main level.

Table (3) displays the results of ADF test. From the results it can be concluded that EOF, is stationary at level, this with confident 95%, as p-value at their level less than 5%, while the all-other variables are not stationary at their level but become difference when taking the first difference this with confident 95%. [2]

*Graph (1):* Line plot for the variables of the study

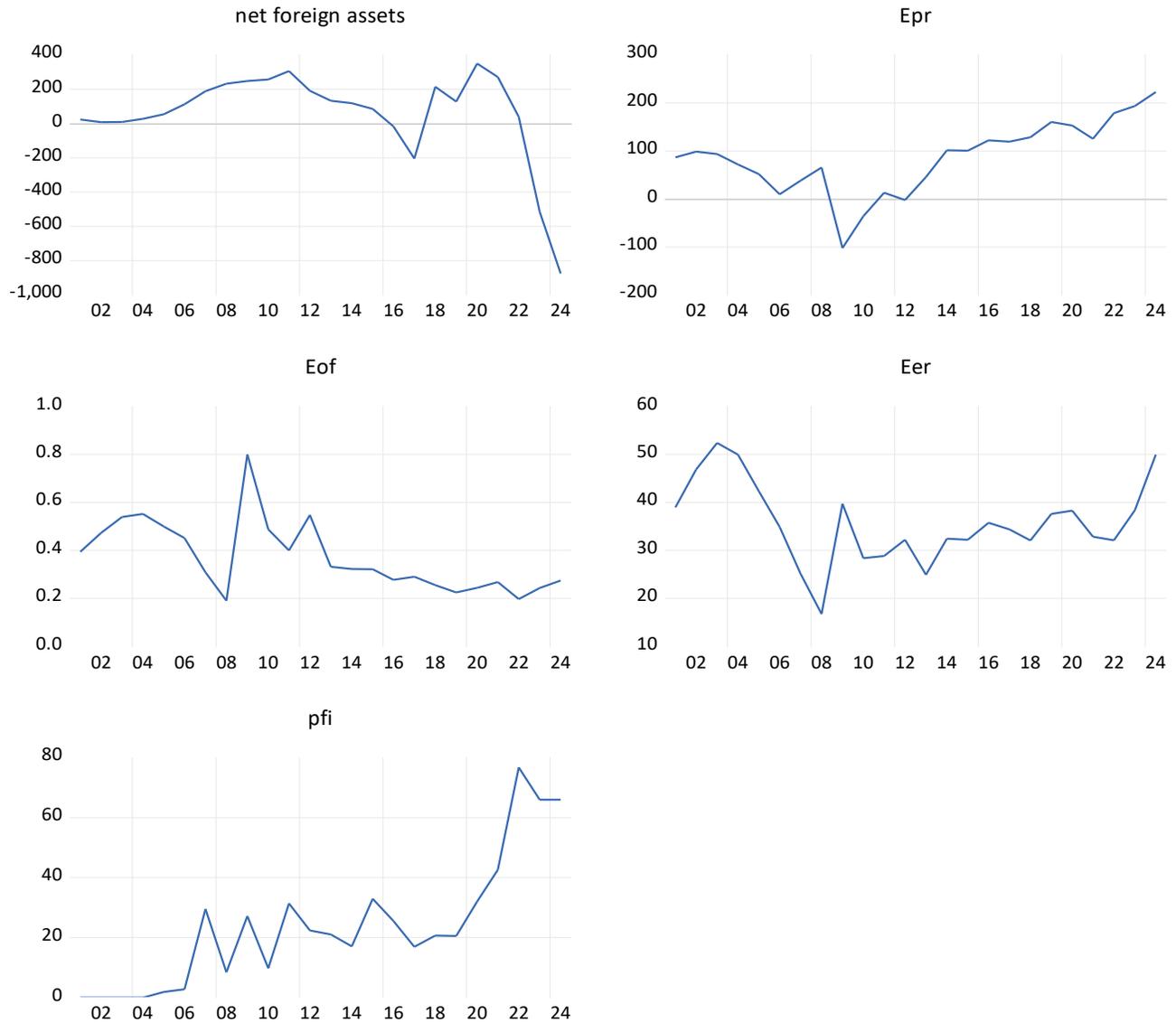

*Table (3):* Augmented Dickey-Fuller (ADF) test for unit root variable

| Variable | ADF | p-value |
|---|---|---|
| Net foreign assets | -2.1485 | 0.2293 |

---

[2] The ADF tests include an intercept. The appropriate lag lengths were selected according to the Schwartz Bayesian criterion, also p-value are calculated using MacKinnon (1996) one-sided p-values. All results are obtained from EViews 12 package.

| | | | |
|---|---|---|---|
| Δ Net foreign assets | | -3.9659 | 0.004*** |
| EER | | -2.135 | 0.2339 |
| Δ EER | | -5.182 | 0.0004*** |
| EOF | | -3.0047 | 0.0492 |
| EPR | | -0.962 | 0.749*** |
| Δ EPR | | -5.577 | 0.0002*** |
| PFI | | -0.0855 | 0.9396 |
| ΔPFI | | -7.731 | 0.0000 |

*10%, **5%, ***1% significance. ADF t-statistic reported.

Table (4): Bounds Test for Cointegration Relationship

| | Critical Value Bounds of the F-Statistic: intercept and no trend (Case II) | | | | | |
|---|---|---|---|---|---|---|
| K | 90% level | | 95% level | | 99% level | |
| 4 | I(0) | I(1) | I(0) | I(1) | I(0) | I(1) |
| | 2.45 | 3.52 | 2.86 | 4.01 | 4.59 | 5.06 |
| | Calculated F-Statistic: | | | | | 2.659 |

Table (5): Estimated the ARDL Model

| Variable | Coefficient | Std. Error | t-Statistic | Prob.* |
|---|---|---|---|---|
| NFA (-1) | 0.889638 | 0.244399 | 3.640107 | 0.0022 |
| EOF | -14.969 | 1.338 | -11.18759342 | 0.0000 |
| EPR | -3.125943 | 0.950817 | -3.287638946 | 0.0000 |
| EER | 11.39837 | 0.49153 | 23.18957134 | 0.0000 |
| PFI | -5.119003 | 0.021474 | -238.381438 | 0.0000 |
| C | 409.5775 | 255.8266 | 1.600996534 | 0.1289 |

Table (6): *Model Criteria/Goodness of Fit*

| | | | |
|---|---|---|---|
| R-squared | 0.743288 | Mean dependent var | 58.56478 |
| Adjusted R-squared | 0.647020 | S.D. dependent var | 274.0415 |
| S.E. of regression | 162.8138 | Akaike info criterion | 13.26888 |
| Sum squared resid | 424133.2 | Schwarz criterion | 13.61447 |
| Log likelihood | -145.5921 | Hannan-Quinn criter. | 13.35579 |
| F-statistic | 7.721094 | Durbin-Watson stat | 2.007264 |
| Prob(F-statistic) | 0.000506 | | |

Table (7): *Autocorrelation test.*

| | AC | PAC | Q-Stat | Prob* |
|---|---|---|---|---|
| 1 | -0.318 | -0.318 | 2.6473 | 0.104 |
| 2 | 0.214 | 0.126 | 3.9034 | 0.142 |
| 3 | -0.224 | -0.141 | 5.3498 | 0.148 |
| 4 | 0.081 | -0.047 | 5.5480 | 0.236 |
| 5 | -0.318 | -0.299 | 8.7803 | 0.118 |
| 6 | -0.014 | -0.258 | 8.7869 | 0.186 |
| 7 | 0.176 | 0.216 | 9.8949 | 0.195 |

| 8  | -0.186 | -0.205 | 11.219 | 0.190 |
| 9  | 0.078  | -0.173 | 11.467 | 0.245 |
| 10 | -0.132 | -0.209 | 12.236 | 0.270 |
| 11 | 0.215  | 0.015  | 14.454 | 0.209 |
| 12 | -0.174 | 0.022  | 16.030 | 0.190 |

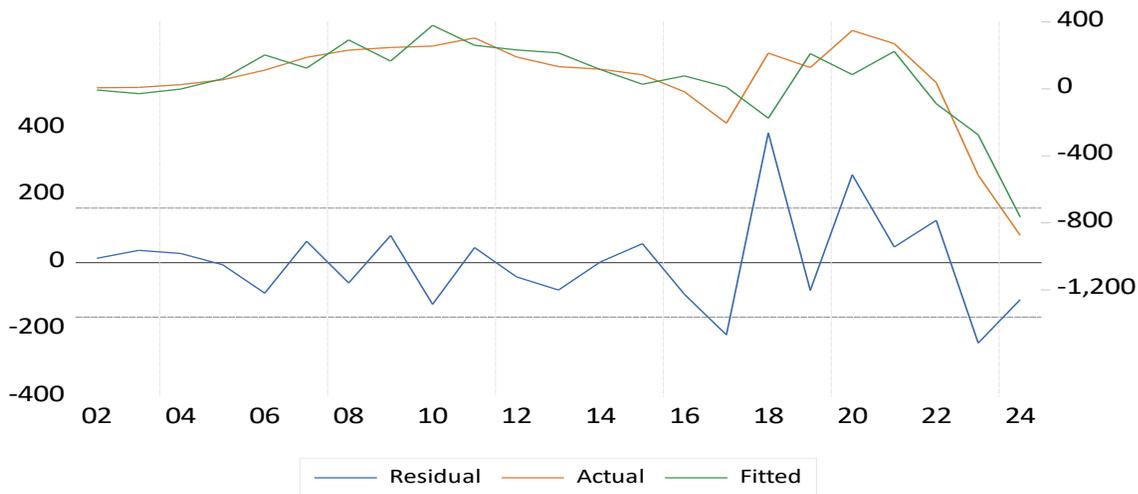

Graph (2): Actual, fitted, residual plot.

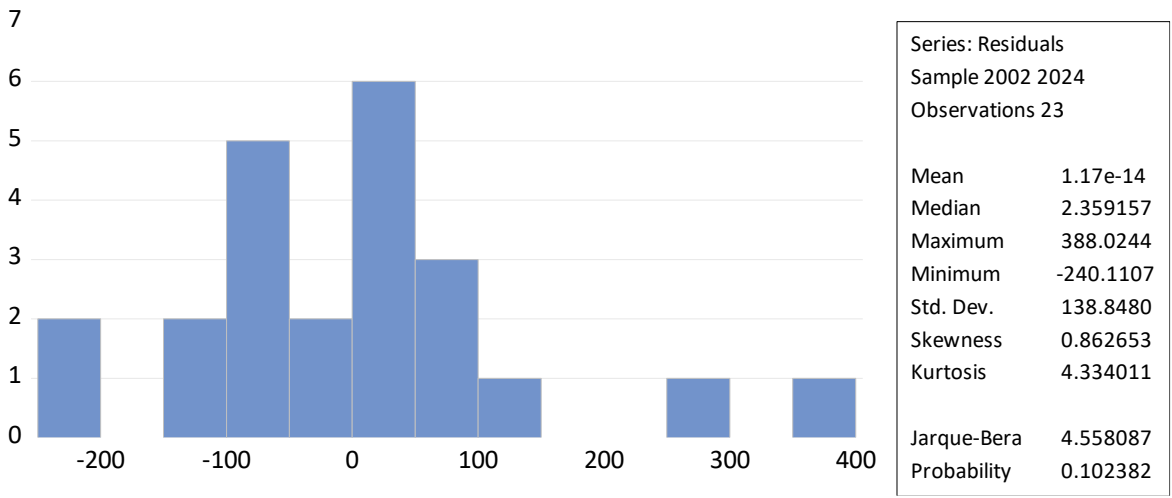

Graph (3): normality of residuals.

Table (8): *Heteroskedasticity Test: Breusch-Pagan-Godfrey*

| Heteroskedasticity Test: Breusch-Pagan-Godfrey | | | |
|---|---|---|---|
| Null hypothesis: Homoskedasticity | | | |
| F-statistic | 1.939898 | Prob. F(6,16) | 0.1356 |

| Obs*R-squared | 9.685667 | Prob. Chi-Square (6) | 0.1385 |
| Scaled explained SS | 7.813595 | Prob. Chi-Square (6) | 0.2521 |

Table (9): The forecasted values for all variables.

|  | EER_F | EOF_F | EPR_F | PFI_F | NET_FOREIGF |
|---|---|---|---|---|---|
| 2025 | 44.08 | 0.21 | 199.99 | 70.33 | -973.25 |
| 2026 | 40.79 | 0.19 | 181.86 | 72.61 | -1055.90 |
| 2027 | 38.93 | 0.18 | 167.12 | 75.88 | -1096.71 |
| 2028 | 37.89 | 0.17 | 155.15 | 78.68 | -1097.23 |
| 2029 | 37.30 | 0.15 | 145.41 | 81.70 | -1064.88 |

Graph (4): The forecasted values of the net foreign assets

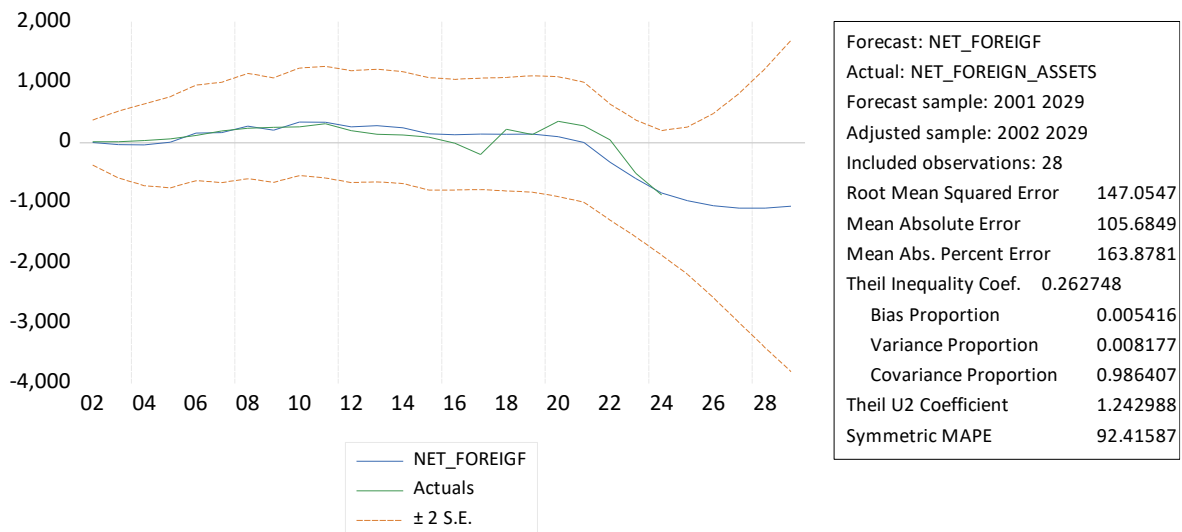

### References:

1. Ahmed, Y. et al, (2020). "Egypt's Engagement with the BRICS: Alternative Development Cooperation Initiative." Journal of Agricultural Economics and Social Sciences 11.11: 825-832.
2. Alabdulwahab, S., Abou-Zaid, A., (November 2023) "Sources of real exchange rate fluctuations in Egypt" Review of Economics and Political Science DOI: 10.1108/REPS-05-2022-0032.
3. Ashraf, K. (2024). Trade Liberalization with BRICS: A CGE Model of Egypt [Master's Thesis, the American University in Cairo]. AUC Knowledge Fountain. https://fount.aucegypt.edu/etds/2241.

4. Aydın, L, (2016) "Intra-BRICS Trade Opening and Its Implications for Carbon Emissions: A General Equilibrium Approach". J. Econ. Dev. Stud, 4 (2), 207–218. https://doi. org/ 10.15640/jeds.v4n2a16.
5. Azevedo, D., et al, (2024) "An Evolving BRICS And the Shitting World Order", international trade,BCG.
6. Bassiouny, A., & Tooma, E. (2018). Extracting shadow exchange rates and foreign exchange premia during currency crises: an example from Egypt. Applied Economics Letters, 26(1), 32–36. https://doi.org/10.1080/13504851.2018.1433291
7. Boshkov, T., (2019) Egypt Currency Crisis: Analysis of The Causes International Journal of Information, Business and Management, Vol. 11, No.1, ISSN 2076-9202.
8. Brahim, M., Nefzi, N. and Sambo, H. (2018), "Do remittances impact the equilibrium real exchange rate? The case of MENA Countries", Revue d'economie du developpement, Vol. 26, pp. 65-119.
9. Branco R., (May 2015)" The BRICS: some historical experiences, growth challenges and opportunities" Center for Growth and Economic Development.
10. Castel, V., Mejia. P, Kolster. J., (2011)" The BRICs in North Africa: Changing the Name of the Game?" AFDB, First Annual Quarter 2011.
11. Correa, Luiz et al, (2014). "5 th BRICS Trade and Economic Research Network (TERN) meeting: the impact of mega agreements on BRICS." Center for Global Trade and Investment.
12. Elshahawany, D. (2022). The Impact of Exchange Rate Volatility on Economic Growth in Egypt, Journal of Business Research- Faculty of Commerce, Zagazig University, Forty-four - Issue Three.
13. F. Jawadi Sushanta K. Mallick Ricardo M. Sousa (2014) "Fiscal Policy in the BRICs", studies in Nonlinear Dynamics and Econometrics · April 2014.
forthcoming (Washington: International Monetary Fund).
14. Fredj J, Sushanta K., Ricardo M. Sousa .,(2016)"Fiscal and monetary policies in the BRICS: A panel VAR approach", Economic Modelling Volume 58, November 2016, Pages 535-542.
15. Greene, R., (December2023)" The Difficult Realities of the BRICS' Dedollarization Efforts— and the Renminbi's Role" Carnegie Endowment for International Peace.
16. Hongmei Li, Ruizhe Xu., (2023)" Impact of fiscal policies and natural resources on ecological sustainability of BRICS region: Moderating role of green innovation and ecological governance", Resources Policy 85,103999.

17. Hooijmaaijers, B. (2021). China, the BRICS, and the limitations of reshaping global economic governance, The Pacific Review, VOL. 34, NO. 1, 29–55.
18. INTERNATIONAL MONETARY FUND, (2011)" New Growth Drivers for Low-Income Countries: The Role of BRICs", Prepared by the Strategy, Policy, and Review Department.
19. Keefe, H.G. (2014), The Impact of Remittance Inflows on Exchange Rate Volatility: The Importance of Dollarization and Development, Department of Economics, Fordham University, New York.
20. Larionova, M., (2016)" Russia's 2015 BRICS Presidency: Models of Engagement with International Organizations, International Organisations Research Journal · DOI: 10.17323/1996-7845-2016-02-113.
21. Larionova. M., Shelepov., A, (January 2016)" Explaining G20 and BRICS compliance", International Organisations Research Journal ·
22. Marc J, Falkenberg D, Macsai G, (2024) "Expansion of BRICS: A quest for greater global influence? European Parliamentary Research, PE 760.368.
23. Mlachila, M. and M. Takebe, 2011, "FDI from BRICs to LICs", IMF Working Paper,
24. Mpofu, T.R. (2016), "The determinants of exchange rate volatility in South Africa", Economic Research Southern Africa, Working Paper No. 604. Cape Town, available at: https://www.econrsa.org/ publications/working-papers/determinants-exchange-rate-volatility-south-africa.
25. Mwase, N. and Y. Yang, 2011 "Philosophies of BRIC Development Flows and Implications for LICs," IMF Working Paper, forthcoming, (Washington: International Monetary Fund).
26. Mwase, N., 2011, "Determinants of development financing flows from BRICs to LICs," IMF Working Paper, forthcoming (Washington: International Monetary Fund).
27. Mwase, N., Yang, Y. (March 2012) "BRICs' Philosophies for Development Financing and Their Implications for LICs", IMF Working Paper, Strategy, Policy, and Review Department International Monetary Fund, WP/12/74.
28. O'Neill. J., (November 23), 2007, BRICS AND BEYOND, Goldman Sachs Global Economics Group.
29. Ozcelebi, O. (2019), "Assessment of asymmetric effects on exchange market pressure: empirical evidence from emerging countries", The North American Journal of Economics and Finance, Vol. 48, pp. 498-513, doi: 10.1016/j.najef.2019.03.016.
30. Samake, I. and Y. Yang, 2011, "BRIC Spillovers to LICs," IMF Working Paper, forthcoming (Washington: International Monetary Fund).
28

31. Sebastian, R., 2008, "China-Africa Investments: An Analysis of China's Investments in Africa," China Vest (September 17), pp.
32. Shelepov., A, (January 2015)" BRICS and international institutions: Models of engagement in global Governance", INTERNATIONAL ORGANISATIONS RESEARCH JOURNAL. Vol. 10. No 4.
33. Zhao H., Lesage D. (2020) Explaining BRICS Outreach: Motivations and Institutionalization. International Organisations Research Journal, vol. 15, no 2, pp. 68–91 (in English). DOI: 10.17323/1996-7845-2020-02-05.